\documentclass[prb,preprint]{revtex4-1} 

\usepackage{amsmath,amsfonts,amssymb,mathrsfs}

\newcommand\del{\delta}
\newcommand\ep{\varepsilon}

\newcommand\x{\xi}

\renewcommand\t{\tau}

\renewcommand\j{\psi}
\renewcommand\o{\omega}

\newcommand\F{\Phi}
\newcommand\scr{Schr\"odinger }
\newcommand{\per}{Perelomov }

\newcommand{\U}{\mathscr{U}}

\renewcommand\d{\partial}

\newcommand{\lan}{\langle}
\newcommand{\ran}{\rangle}

\begin{document}


\title{Airy wavepackets are Perelomov coherent states\footnote{This paper is dedicated to the memory of late Prof. E. C. G. Sudarshan, whose works have been a source of inspiration for the author.}}

\author{Vivek M. Vyas}
\email{vivekv@rri.res.in} 
\affiliation{Theoretical Physics Group, Raman Research Institute, Sadashivnagar, Bengaluru 560 080, INDIA}

\date{\today}

\begin{abstract}
Accelerating non-spreading wavepackets in nonrelativistic free particle system, with probability distribution having an Airy function profile, were discovered by Berry and Balazs (1979), and have been subsequently realised in several optical experiments. It is shown that these wavepackets are actually Perelomov coherent states. It is found that the Galilean invariance of the \scr equation plays a key role in making these states unique and giving rise to their unusual propagation properties.
\end{abstract}

\maketitle 

\section{Introduction}

In a remarkable development, Berry and Balazs \cite{berry} showed that the wavefunction $\j(x,t)$ of a nonrelativistic free particle system, which evolves as per the \scr equation:
\begin{align} \label{eq1}
	- \frac{\hbar^2}{2 m} \frac{\d^2 \j }{\d x^2} = i \hbar \frac{\d \j }{\d t}, 
\end{align}
admits a solution having an Airy function profile:
\begin{align} \label{sol1}
	\psi(x,t=0) = \text{Ai} \left(\frac{B x}{\hbar^{2/3}}\right),
\end{align}
which uniformly accelerates without spreading:
\begin{align} \label{sol2}
	| \j (x,t) |^2 = \text{Ai}^2 \left(\frac{B}{\hbar^{2/3}} \left( x - \frac{B^3 t^2}{4 m^2}\right)\right).
\end{align}
It was shown that the acceleration of these wavepackets is not in conflict with the Ehrenfest theorem, since these states are not square-integrable, giving rise to ill-defined averages $\lan x \ran$ and $\lan x^2 \ran$. Infact it was argued that $| \j (x,t) |^2$ does not represent probability density of a single particle.

Airy wavepackets have been a subject of careful investigation from various aspects.\cite{green, rau} In the last several years, the optical counter parts of these Airy wavepackets have been a subject of intense study. It is well known that, the Helmholtz equation governing the dynamics of the electric field envelope of a plane polarised optical beam, becomes identical to free \scr equation, in the paraxial approximation.\cite{good} This identification naturally provides a route for unhindered exchange of concepts and results between these two different physical systems, which was exploited by Siviloglou and Christodoulides, to show the possibility of realisation of the optical beam counterpart of non-spreading accelerating Airy wavepacket.\cite{Sivi} In 2007, such optical Airy beams were experimentally realised by Siviloglou \emph{et. al.}, and since then they have been keenly studied.\cite{siv} Apart from being non-diffracting beam, which makes them useful in their own right,\cite{dholakia} these beams also display self-healing,\cite{self} possess a nontrivial orbital angular momentum behaviour \cite{alf} and polarisation property.\cite{rpg} These unique properties makes such beams suitable for a number of applications, generation of plasma in a dielectric media in a controlled manner \cite{poly} and optical manipulation of dielectric microparticles \cite{bau} are two notable ones amongst the others.    


In an introductory quantum mechanics course, one is introduced to the \scr equation obeyed by the wave function of the system and its probabilistic interpretation. While discussing the quantum mechanical harmonic oscillator, one comes across the celebrated \emph{coherent (wavepacket) states} $\phi(x,t)$, which evolve in time in a manner so as to preserve its shape: 
\begin{align}
	|\phi(x,t)|^{2} = |\phi(x - f(t),0)|^{2}.
\end{align}
Interestingly, one learns that $f(t)$ solves the classical equation of motion for the harmonic oscillator, and as a result the peak of the probability density follows a classical trajectory for these states.\cite{gott} Incidentally from Eq. (\ref{sol2}) one sees that the Airy wavepackets also evolve in time similar to these coherent states, preserving their shape:
\begin{align}
	| \j (x,t) |^2 = | \j (x - g(t),0) |^2,
\end{align} 
where $g(t) = \frac{B^3 t^2}{4 m^2}$. This observation leads to a natural question: whether Airy wavepackets have any connection with the harmonic oscillator coherent states ? To the best of the author's knowledge this natural question has not been yet answered. The goal of this paper is to bring to fore an interesting relation between these two wavepackets, and show that the non-spreading Airy wavepackets are actually generalised coherent states of Perelomov kind.

\section{Harmonic oscillator coherent state and its generalisation}
The coherent states were first studied by Schr\"odinger, and came into prominence from the works of Sudarshan and Glauber in quantum optics.\cite{klauder, gazeau} It is worth recalling that, in terms of annihilation operator: $\hat{a} = \sqrt{\frac{m \o}{2 \hbar}} \hat{x} + i \frac{1}{\sqrt{2 \hbar m \o}} \hat{p}$, and creation operator $\hat{a}^\dagger$, such that $[\hat{a}, \hat{a}^\dagger] = \hat{I}$; the harmonic oscillator Hamiltonian reads: $\hat{H} = \hbar \o \left(\hat{a}^\dagger \hat{a} + \frac{1}{2} \hat{I}\right)$. The coherent state in this system can be constructed from the ground state $|0\ran$, which is annihilated by $\hat{a}$: $\hat{a}|0\ran = 0$, from the application of a unitary operator, which is constructed using $\hat{a}^{\dagger}$ and $\hat{I}$, on it:\cite{gott}
\begin{align}
	|\alpha \ran = \exp \left( \alpha \hat{a}^{\dagger} - \frac{|\alpha|^2}{2} \hat{I} \right) | 0 \ran. 
\end{align}  
There are several properties that make this family of coherent states special.\cite{klauder, gazeau} An important property is its temporal stability, that is, a coherent state is stable under time evolution, which only changes the value of $\alpha$: $e^{- i \frac{\hat{H}t}{\hbar}} | \alpha \ran = | \alpha e^{- i \o t} \ran$. One of the most important property possessed by these states is that the functional form of the probability density remains the same under time evolution, except that the peak of the function performs periodic classical motion around the origin:
\begin{equation*}
	P_\alpha (x,t) = \frac{1}{x_0 \sqrt{\pi}} \exp \left( - \frac{1}{x^2_0} ( x - \lan x (t) \ran_\alpha )^2 \right),
\end{equation*}
where $\lan x (t) \ran_\alpha = \sqrt{2} x_0 |\alpha| \cos (\o t)$, and $x^2_0 = \frac{\hbar}{m \o}$. This shows that this coherent wave packet does not disperse or distort as it evolves.

One naturally wonders whether there exists other quantum system which possess such coherent states or not. It turns out that there exist coherent states in several other quantum systems, but they do not obey all the properties possessed by the harmonic oscillator coherent states. In this scenario, one has to contend working with different generalisations of harmonic oscillator coherent states, each of them possessing distinct trait and character. One of such well studied generalisation are the so called \emph{Perelomov coherent states},\cite{perel,puri} which possess a rich mathematical structure. In a given system, say if the operators $\hat{X}$, $\hat{Y}$ and $\hat{Z}$ form a closed algebra, so that the commutator of any two operators is equal to the third one (modulo a c-number factor), then the Perelomov coherent states can be straightforwardly constructed in such a case. From the state $| x \ran$, which is an eigenstate of $\hat{X}$: $\hat{X} | x \ran = x | x \ran$, the coherent state $|a,b \ran$ can be constructed as:\cite{fn1}
\begin{align}
	| a, b \ran = \exp \left(a \hat{Y} + b \hat{Z} \right) | x \ran. 
\end{align} 
Analogously one can construct coherent states from the eigenstates of other operators as well, which would provide with a different family of coherent states.\cite{fn2} The initial state, which ought to be an eigenstate of an operator which forms the algebra, is often called the fiducial state in the literature.  One notes that, in essence, the Perelomov definition of coherent states aims at giving a group theoretic generalisation of harmonic oscillator coherent states. In certain cases, it has been shown that these Perelomov coherent states show classical behaviour, and saturate the Heisenberg uncertainty relation as well.\cite{perel}

\section{Galilean invariance of \scr equation}

It is a well known fact that, the Newton's laws of motion remain the same for any two observers moving  inertially with respect to one another. This is a manifestation of (Galilean) principle of relativity. This principle also holds in nonrelativistic quantum mechanics, where it manifests as the \emph{form invariance} of the \scr equation for the two observers. An observer observing a particle of mass $m$ moving in one dimensional space, finds that its motion is described by the wavefunction $\j(x,t)$, which solves the \scr equation:
\begin{align}
	- \frac{\hbar^2}{2 m} \frac{\d^2 \j }{\d x^2} = i \hbar \frac{\d \j }{\d t}. 
\end{align}
The same motion is observed by some other observer, who is moving relative to the former with a constant velocity $v$, using coordinates $x' = x - vt$ and $t'=t$. He finds that the motion is being described by the wavefunction $\j'(x',t')$ which solves the \scr equation:
\begin{align}
	- \frac{\hbar^2}{2 m} \frac{\d^2 \j' }{\d {x'}^2} = i \hbar \frac{\d \j' }{\d t'}, 
\end{align}
albeit in his coordinate frame. It can be easily checked that the two wavefunctions $\j$ and $\j'$ are connected via a unitary transformation:\cite{gott}
\begin{align} \label{ut1}
	\j'(x',t') = \exp \left[\frac{i}{\hbar}\left(\frac{m v^2 t}{2} - m v x \right)\right] \j(x,t),
\end{align}
which does not alter their normalisations. This unitary transformation which connects the two inertial observers can be simply seen to be expressed in terms of the operator:
\begin{equation} \label{K}
	\hat{K}(t)= t \hat{p} - m \hat{x},
\end{equation}
called the \emph{Galilean boost (generator)}.\cite{gott} This can be seen by writing it as: 
\begin{align} 
	\lan x' | \j'(t')\ran &= e^{-i \frac{ m v^2 t}{2 \hbar} } \lan x' |  e^{ - i \frac{v m }{\hbar} \hat{x}} e^{ i \frac{v t }{\hbar} \hat{p}} | \j(t) \ran \\ \label{ut2} &= \lan x' | e^{i \frac{v}{\hbar} \hat{K}(t)} | \j(t) \ran,
\end{align}
($|x\ran$ is a position eigenstate with eigenvalue $x$) which shows that any given state $|\j(t)\ran$ for the former observer corresponds to the transformed state (called the \emph{boosted} state): 
\begin{align} \label{bt}
	|\j'(t')\ran = e^{i \frac{v}{\hbar} \hat{K}(t)} | \j(t) \ran
\end{align}
to the latter. Here $| x' \ran = | x - v t \ran = e^{i v t \hat{p}/\hbar} |x \ran$. Interestingly both the observers agree that the average of $\hat{K}(t)$ itself is a constant of motion: $\frac{d}{dt} \lan \j (t)| \hat{K}(t)|\j(t) \ran = 0$ and  $\frac{d}{dt'} \lan \j' (t')| \hat{K}(t')|\j'(t') \ran = 0$. This is a consequence of \scr equation and can be easily derived using relations: $m \frac{d \lan x \ran}{dt}=\lan p \ran$ and $\frac{d \lan p \ran}{dt} = 0$.

\section{Accelerating free particle coherent states}

In a given system the Hamiltonian,  which in our case is $\hat{H}=\frac{\hat{p}^2}{2 m}$, generates the time evolution. The stationary states of the \scr equation, being the eigenstates of Hamiltonian, evolve trivially under time evolution. One wonders whether there exists states which get trivially transformed under a boost transformation, given by Eq. (\ref{bt}). Using the analogy one finds that such states would be the eigenstates of $\hat{K}$:
\begin{equation}
	\hat{K}(t) | \x;t \ran = \x | \x;t \ran,
\end{equation} 
with real eigenvalue $\x$. Note that the set of operators \{ $\hat{I}$, $\hat{x}$, $\hat{p}$, $\frac{\hat{p}^2}{2}$ \} with the commutation relations: 
\begin{align}
	[\hat{x} , \hat{p}] = i \hbar \hat{I}, \quad [\hat{x} , \frac{\hat{p}^2}{2} ] = i \hbar \hat{p}, \quad
	[\hat{p} , \frac{\hat{p}^2}{2} ] = 0, 
\end{align} 
constitute a closed algebra. This algebra can be thought of as a generalisation of harmonic oscillator algebra of \{$ \hat{a}, \hat{a}^\dagger, \hat{I}$ \}, which can also be expressed using the operators \{$\hat{I}$, $\hat{x}$, $\hat{p}$\}.

Considering $|\F\ran = |x=0;t=0\ran$ as the fiducial state,\cite{fn3} one can now construct Perelomov coherent states from it by the application of unitary operators constructed using $\hat{p}$ and $\frac{\hat{p}^2}{2}$. Surprisingly one finds that the coherent state so generated is actually the state $| \x;t \ran$:
\begin{align}
	| \x;t\ran = \exp \left( - \frac{i t}{2 \hbar m}\hat{p}^2 + \frac{i}{\hbar}\frac{\x}{m} \hat{p} \right) |\F\ran,
\end{align}
for any real $\x$. 
Being eigenstates of a Hermitean operator $\hat{K}$, these states form an orthonormal complete basis set:
\begin{align}
	\int\limits_{-\infty}^{\infty} d\x | \x;t\ran \lan \x;t | = 1 \quad \text{and} \quad \lan \x;t| \x' ;t\ran = \del(\x-\x').
\end{align}
These relations can be explicitly checked using their $x$-representation: 
\begin{align}
	\j_{\x}(x,t) = \lan x | \x;t\ran = \frac{1}{\sqrt{2 \pi \hbar |t|}} \exp \left( \frac{i}{\hbar} \left( \frac{m x^2}{2 t} + \frac{\x x}{t} \right)\right).
\end{align}

Note that the set of operators \{$\hat{I}$, $\hat{x}$, $\hat{p}$, $\frac{\hat{p}^2}{2}$, $\frac{\hat{p}^3}{6}$\}, with nontrivial commutation relations:
\begin{align}
	[\hat{x} , \hat{p}] = i \hbar \hat{I}, \quad [\hat{x} , \frac{\hat{p}^2}{2} ] = i \hbar \hat{p}, \quad
	[\hat{x} , \frac{\hat{p}^3}{6} ] = i \hbar \frac{\hat{p}^2}{2},
\end{align} 
form a closed algebra. This motivates one to construct a generalisation of coherent state $| \x;t \ran$, from the fiducial state $|\F\ran = |x=0;t=0\ran$, albeit using the operators \{$\hat{p}$, $\frac{\hat{p}^2}{2}$, $\frac{\hat{p}^3}{6}$\}. Such a coherent state $|\ep,\x;t\ran$ is given by:
\begin{align} \label{ut}
	|\ep,\x;t\ran &= \U(\ep,t,\x) |\F\ran \\ \label{utt}
	& =\exp \left(- \frac{i \ep}{6 \hbar m^2} \hat{p}^3 - \frac{i t }{2 \hbar m}{\hat{p}}^2 + \frac{i \x }{\hbar m} \hat{p}\right)  |\F\ran,
\end{align}
for any real $\ep$ and $\x$. Interestingly it turns out that this state is an eigenstate of $\hat{K}(t) + \ep \hat{H}$:
\begin{align} \label{gcs}
	\left(\hat{K}(t) + \ep \hat{H} \right)|\ep,\x;t\ran = \x |\ep,\x;t\ran.
\end{align} 
The coherent states $|\x,t\ran$ are a special case of these general coherent states corresponding to $\ep = 0$. It immediately follows from Eq. (\ref{gcs})  that for the same value of $\ep$, the states $|\ep,\x;t\ran$ with different $\x$ span the Hilbert space and provide with a complete orthonormal basis set:
\begin{align}
	\int\limits_{-\infty}^{\infty} d\x |\ep, \x;t\ran \lan \ep,\x;t | = 1 \quad \text{and} \quad \lan \ep,\x;t|\ep, \x' ;t\ran = \del(\x-\x').
\end{align} 
The states with same $\x$ but different $\ep$ are however \emph{not} orthogonal:
\begin{align}
	\lan \ep,\x;t|\ep', \x ;t\ran = \left(\frac{2 \hbar m^2}{\ep - \ep'}\right)^{\frac{1}{3}} \text{Ai}(0), 
\end{align}
which can be easily deduced by working with the $p$-representation $\tilde{\j}_{\ep,\x}(p,t)$ of state $|\ep, \x ;t\ran$:
\begin{align} \label{prep}
	\tilde{\j}_{\ep,\x}(p,t) = \lan p |\ep, \x ;t\ran = \frac{1}{2 \pi \hbar \sqrt{m}} e^{\frac{i}{\hbar}(\frac{p \x}{m} - \frac{ t p^2 }{2 m} - \frac{ \ep p^3 }{6 m^2})}.
\end{align}
 
The fact that $|\ep, \x;t\ran$ is an eigenstate of $\hat{K}(t) + \ep \hat{H}$, gives it a unique dynamical behaviour. 
Using the Zassenhaus formula:
\begin{align}
	e^{t (X + Y)} = e^{t X} e^{t Y} e^{- \frac{t^2}{2} [X,Y]} e^{\frac{t^3}{6} \left(2 [Y,[X,Y]] + [X,[X,Y]] \right)} \dots
\end{align}
one can decompose exponential of $\hat{K}(t) + \ep \hat{H}$ as:
\begin{align} \label{zas}
	e^{\frac{i v}{\hbar}(\hat{K}(t) + \ep \hat{H})} = e^{-\frac{i m \ep v^3}{3 \hbar}} e^{\frac{i v \ep}{\hbar}\hat{H}} e^{\frac{i v}{\hbar}\hat{K}(t)}  e^{\frac{i v^2 \ep}{2 \hbar}\hat{p}}.
\end{align} 
Noting that the product $\ep v$ has dimensions of time, and denoting it as $\t (= \ep v)$, this equation on state $|\ep, \x ;t\ran$ at $t=0$ leads to a very interesting relation:
\begin{align} \label{rel}
	e^{-\frac{i \t }{\hbar} \hat{H}} |\ep, \x ;0\ran =  \left( e^{-\frac{i \t \x}{\hbar \ep}} e^{-\frac{i m \t^3}{3 \hbar \ep^3}} \right)  e^{\frac{i \t }{\hbar \ep}\hat{K}(0)} e^{\frac{i \t^2 }{2 \hbar \ep}\hat{p}} |\ep, \x ;0\ran.
\end{align}
It essentially shows that effect of time evolution operator on such a state is same as that of action of spatial translation operator, albeit with time dependent translation parameter $\frac{\t^2 }{2 \ep}$; followed by a Galilean boost, albeit with a time dependent velocity $\frac{\t}{\ep}$. A little reflection will convince the reader that these two operations actually lead to a constant acceleration $\frac{1}{\ep}$. This shows the unique feature of dynamics of these coherent states:  time evolution is (modulo a time dependent phase) is same as being transformed with a constant acceleration. 

Above inference about the constant accelerating nature of the coherent states was done by inspecting the effect of time evolution operator on them. However the reader might wonder whether this acceleration physically manifests in any aspect or whether is just a mathematical artefact. This can be best addressed by working in $x$-representation, were $\j_{\ep,\x}(x,t) = \lan x | \ep, \x;t\ran $ stand for the wavefunction corresponding to state $|\ep, \x;t\ran$. The relation (\ref{rel}) in this representation reads:
\begin{align} \label{simple}
	\j_{\ep,\x}(x,\t) = e^{-\frac{i \t \x}{\hbar \ep}} e^{- \frac{i}{\hbar} (\frac{m \t^3}{3 \ep^2} + \frac{m x \t}{\ep})} \j_{\ep,\x}(x + \frac{\t^2}{2 \ep},0).
\end{align} 
This elegant relation depicts that the shape of probability density for such (wave packet) states does not change under time evolution, and they accelerate with a constant acceleration $\frac{1}{\ep}$:
\begin{align}  
	|\j_{\ep,\x}(x,\t)|^2 = |\j_{\ep,\x} (x + \frac{\t^2}{2 \ep},0)|^2.
\end{align} 
This is a clear proof of the non-spreading nature of these wave packet states: the form of the density remains the same, the time evolution only transports it quadratically as a function of time.    
On finding the explicit form of $\j_{\ep,\x}(x,t)$:
\begin{align} \nonumber
\lan x | \ep, \x;t\ran = \j_{\ep,\x}(x,t) = &\frac{1}{\sqrt{m \hbar^2}} \left(\frac{2 \hbar m^2}{\ep}\right)^{\frac{1}{3}} \times 
\exp \left( -\frac{i}{\hbar} \frac{(\x + m x)t}{\ep} - \frac{i}{\hbar} \frac{m t^3}{3 \ep^2}\right) \\& \times
\text{Ai} \left(-\frac{1}{\hbar} \left(\frac{2 \hbar m^2}{\ep}\right)^{\frac{1}{3}} \left( x + \frac{\x}{m} + \frac{t^2}{2 \ep} \right) \right)
\label{solnf},
\end{align}
one surprisingly discovers that, these Perelomov coherent states are actually the Airy wavepackets encountered in Eq. (\ref{sol1}), found by Berry and Balazs.\cite{berry} This remarkable identification of Airy wavepackets as Perelomov coherent states is the main result of this work. It clarifies the origin of the accelerating motion and the non-spreading nature of the Airy wavepacket, both of which are consequences arising out of its definition (\ref{utt}) as a Perelomov coherent state.

It is worth mentioning that unlike harmonic oscillator coherent states, the peak of the probability density of these coherent states {does not} traverse a classical trajectory, which in this system is the one without any acceleration, but rather traverses along a caustic of classical trajectories, as elucidated by Berry and Balazs,\cite{berry} and Berry.\cite{berrys} It is worth emphasizing that, from this treatment it becomes apparent that the Galilean invariance of free \scr equation has played a vital role in realisation of these coherent states in the system. The behaviour of \scr equation under Galilean transformation and its generalisation, in this context, from the view point of gravitation has been studied in detail by Greenberger.\cite{green}   

In the presence of a potential, the Galilean and translational invariance of the \scr equation is lost, and as a result such non-spreading accelerating coherent states can not be constructed in any other system with a potential.  

From the above discussion the physical significance of the parameter $\ep$  
and the nature of unitary transformation (\ref{ut}) which generates it, becomes clear. The action of $\U(\delta,0,0)$ on a state accelerating with acceleration $\frac{1}{\kappa}$ transforms it to the one with acceleration $\frac{1}{\kappa + \delta}$. 

The state of zero acceleration can be obtained by taking the limit $\ep \rightarrow \infty$ with a fixed finite value of $\x$. The relation (\ref{rel}) in this limit yields:
\begin{align} e^{-\frac{i \t }{\hbar} \hat{H}} |\ep \rightarrow \infty , \x ;0\ran = |\ep \rightarrow \infty , \x ;0\ran
\end{align}
which shows that the state $|\ep \rightarrow \infty , \x ; \t \ran$ is actually the ground state of the system with zero energy $|p=0\ran$, since it is the only state that does not evolve in time. On the other hand, in the limit $\ep \rightarrow 0$, the state $|\ep, \x ;0\ran$ goes over to become the eigenstate of $\hat{K}$. 
Thus the accelerating coherent states $|\ep , \x ; \t \ran$ are seen to interpolate smoothly and continuously from the ground state $|p=0\ran$, which has no acceleration; to $\hat{K}$ eigenstate $|\ep = 0, \x ; \t \ran$.    

\section{Conclusion}

In this paper, it is shown that the accelerating wavepacket solutions of free \scr equation, discovered by Berry and Balazs, are indeed Perelomov coherent states. The harmonic oscillator coherent states are known to arise out of the algebra generated by \{$\hat{I}$, $\hat{x}$, $\hat{p}$\}. By considering a larger algebra, consisting of \{$\hat{I}$, $\hat{x}$, $\hat{p}$, $\frac{\hat{p}^2}{2}$, $\frac{\hat{p}^3}{6}$\}, it is found that the accelerating Airy wavepackets arise out of it as Perelomov coherent states. Moreover it is found that these coherent states solve the eigenvalue problem for the linear combination of boost operator $\hat{K}$ and Hamiltonian $\hat{H}$, which is responsible for their seemingly intriguing non-spreading accelerating nature. This provides one with a representation independent understanding of the origin and the nature of such accelerating coherent states. To the best of the author's knowledge, all the other works concerning origin of these accelerating wavepacket states commit to some or the other representation. Moreover the present treatment is naturally suitable for a realisation and identification of such coherent states in systems involving statistical mixtures and in open systems, wherein the quantum free particle is interacting with a larger reservoir system, while being in or away from thermal equilibrium. It also allows one to straightforwardly construct such coherent states in systems involving quantised fields. Such systems naturally appear while dealing with indefinite and large number of particles, and are often encountered in quantum optics and condensed matter experiments.

\section*{Acknowledgements}
The author thanks Prof. P. K. Panigrahi from IISER Kolkata and Prof. V. Srinivasan of University of Madras for several stimulating discussions and their constant encouragement. The author is indebted to Prof. R. P. Singh of Physical Research Laboratory, Ahmedabad for many discussions on various aspects of Airy beams and for providing financial support in the initial stage of this work. The author acknowledges a useful conversation with Prof. J. Samuel and Dr. D. Roy of Raman Research Institute. The author thanks Prof. Michael Berry for an encouraging and enlightening communication. The author acknowledges constructive comments and suggestions received from the anonymous reviewers, which has helped the author gain a better understanding and has added clarity to the paper.          


\begin{thebibliography}{100}
\bibitem{fn1}{A familiar example of these coherent states is given by angular momentum coherent states, obtained by the action of $SO(3)$ operators on $\hat{L}_z$ eigenstate: $\exp ( i \theta_x \hat{L}_x + i \theta_y \hat{L}_y ) | m \ran$, where $\hat{L}_z | m \ran = m | m \ran$.}
\bibitem{fn2}{What is presented here is the essence of the coherent state construction due to Perelomov. For a mathematically rigorous and nuanced treatment, the reader is referred to Refs. \onlinecite{gazeau} and \onlinecite{perel}.}
\bibitem{fn3}{One can consider any other fiducial state say $|x = x_0; t = t_0 \ran$, without affecting subsequent conclusions about \per coherent states. The state $|x=0;t=0\ran$ is only chosen for simplicity.}
\end{thebibliography}

\end{document}